\documentclass[%
 reprint,
preprintnumbers,
 amsmath,amssymb,
 aps,
 prl,
]{revtex4-1}
\usepackage{graphicx}
\usepackage{dcolumn}
\usepackage{bm}
\usepackage{hyperref}

\usepackage[ngerman,english]{babel}
\usepackage[latin1]{inputenc}

\usepackage[alsoload=synchem]{siunitx}

\usepackage{upgreek}

\begin{document}


\title{Strain Engineering of the Band Gap of HgTe Quantum Wells using Superlattice Virtual Substrates}

\author{Philipp Leubner}
 \email{philipp.leubner@physik.uni-wuerzburg.de}
\author{Lukas Lunczer}%
\author{Christoph Brüne}%
\author{Hartmut Buhmann}%
\author{Laurens W. Molenkamp}%
\affiliation{%
 Experimentelle Physik III, Physikalisches Institut, Universität Würzburg, Am Hubland, D-97074, Würzburg, Germany
}%
 
\date{\today}

\begin{abstract}
The HgTe quantum well (QW) is a well-characterized two-dimensional topological insulator (2D-TI). Its band gap is relatively small (typically on the order of $\SI{10}{\milli\electronvolt}$), which restricts the observation of purely topological conductance to low temperatures. 
Here, we utilize the strain-dependence of the band structure of HgTe QWs to address this limitation. We use $\text{CdTe-Cd}_{0.5}\text{Zn}_{0.5}\text{Te}$ strained-layer superlattices on GaAs as virtual substrates with adjustable lattice constant to control the strain of the QW. 
We present magneto-transport measurements, which demonstrate a transition from a semi-metallic to a 2D-TI regime in wide QWs, when the strain is changed from tensile to compressive. Most notably, we demonstrate a much enhanced energy gap of $\SI{55}{\milli\electronvolt}$ in heavily compressively strained QWs. This value exceeds the highest possible gap on common II-VI substrates by a factor of 2-3, and extends the regime where the topological conductance prevails to much higher temperatures.

\end{abstract}

\pacs{Valid PACS appear here}
\maketitle


The transport properties of molecular-beam epitaxially (MBE) grown HgTe QWs embedded in $\text{Cd}_{0.7}\text{Hg}_{0.3}\text{Te}$ barriers have attracted considerable attention due to the discovery of the quantum-spin-Hall (QSH) effect in these structures \cite{konig2007,roth2009,brune2012}.  The QSH effect is the landmark property of a 2D-TI and is characterized by the presence of a pair of one-dimensional, counter-propagating (``helical'') channels along the edges of the mesa, giving rise to a quantized longitudinal conductance $G_\text{QSH} = e^2\ h^{-1}$ \cite{konig2007}. A prerequisite for the formation of edge channels is a - topologically nontrivial - inverted band structure, as is present in HgTe QWs when the thickness $d_\text{QW}$ exceeds $d_\text{c} = \SI{6.3}{\nano\metre}$ \cite{konig2007}. Inverted HgTe QWs have a relatively small band gap $E_\text{G}$ (typically lower than 15 meV), which can make it difficult to gate homogeneously into the gap over the whole mesa, and also prevents applications at elevated temperatures. Here we present a way to increase $E_\text{G}$ well above the thermal energy at room temperature ($k_\text{B}T = \SI{25}{\milli\electronvolt}$). This is achieved by applying compressive strain to HgTe QWs through coherent growth on virtual substrates with a freely tunable lattice constant.  

The crucial influence of strain on the band structure of HgTe has been demonstrated previously for bulk layers (layer thickness $d > \SI{40}{\nano\metre}$): epitaxy of HgTe on CdTe substrates exerts tensile strain ($\varepsilon =  -\SI{0.3}{\percent}$), which causes a gap-opening of the  $\Gamma_8$ doublet, transforming the bulk semimetal into a three-dimensional topological insulator \cite{brune2011,brune2014}. However, these previous experiments used commercially available MBE quality substrates, limiting the options to $\text{Cd}_{0.96}\text{Zn}_{0.04}\text{Te}$\cite{konig2007,roth2009,brune2012} and CdTe \cite{brune2011,brune2014}\footnote{Alternatively, several $\upmu\text{m}$ thick buffer layers of fully relaxed CdTe on Si or GaAs can be employed}. In both cases, the lattice constant of the substrate material is larger than that of HgTe, resulting in a  tensile strain in the epilayers. Under such conditions, the largest gaps that can be obtained in inverted QWs are $E_\text{G} = \SI{17}{\milli\electronvolt}$ and $\SI{25}{\milli\electronvolt}$ for wells grown on CdTe and $\text{Cd}_{0.96}\text{Zn}_{0.04}\text{Te}$, respectively \cite{PfeufferJeschke2000}.\par

The present work reports on a major progress in this situation. We use $\text{CdTe-Cd}_{0.5}\text{Zn}_{0. 5}\text{Te}$ (001) strained-layer superlattices (SLS) as virtual substrate material for HgTe-based epilayers. These superlattices (grown on a GaAs substrate) provide a straightforward control of the effective lattice constant of the system, and thus the strain in the subsequently grown HgTe layers. The use of SLS rather than (Cd,Zn)Te solid solutions for lattice constant control is necessary because solid solutions suffer from poor crystal quality due to phase separation effects \cite{feldman1986,feldman1987}. We fabricate both tensile ($\varepsilon < 0$) and compressively ($\varepsilon > 0$) strained QWs using coherent epitaxy of (Zn,Cd,Hg)Te - HgTe - (Zn,Cd,Hg)Te heterostructures on virtual substrates. High resolution-X-ray-diffraction (HRXRD) is used to analyze the SLS crystal structure, and to determine the amount of strain introduced in the (Zn,Cd,Hg)Te - HgTe - (Zn,Cd,Hg)Te heterostructure. Magnetic field- and temperature-dependent transport measurements of Hall bar devices reveal that the change from tensile to compressive strain induces a transition from a semimetallic to a 2D-TI system for wide QWs. For thinner QWs under heavy compressive strain, an yet unreported band gap of as high as $E_\text{G} = \SI{55}{\milli\electronvolt}$ is observed. \par

\begin{figure}[b]
\includegraphics{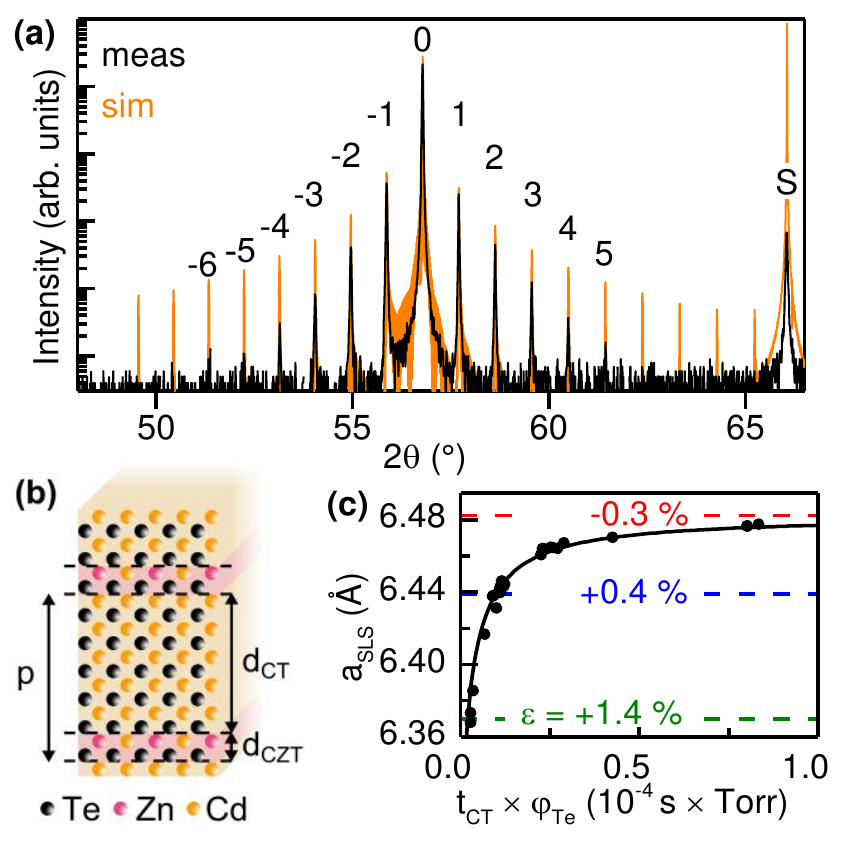}
\caption{\label{fig:1} (a) HRXRD $2\theta - \omega$ scan (black) and simulation (orange) of the (004) reflection of a SLS with superlattice period $p = \SI{109}{\angstrom}$. The (004) reflection from the GaAs substrate is labelled S, the SLS Bragg- and higher order peaks are labelled with 0 and $\pm1,\ \pm2,\ \pm3$... respectively. (b) Sketch of a single SLS period: one ML $\text{Cd}_{0.5}\text{Zn}_{0.5}\text{Te}$ alternating with an adjustable thickness $d_\text{CT}$ of CdTe. (c) Dots: effective lattice constant $a_\text{SLS}$ of a set of SLS with varying product of CdTe growth time and Te beam equivalent pressure $t_\text{CT} \times \varphi_\text{Te}$. The line is a fit to Eq.~\ref{eq:1}. Dashed lines indicate the lattice constants used for samples A ($\varepsilon =  -\SI{0.3}{\percent}$),  B ($\varepsilon =  +\SI{0.4}{\percent}$) and C ($\varepsilon =  +\SI{1.4}{\percent}$).}
\end{figure}

The SLS used in this work are fabricated on GaAs:Si (001) substrates by alternating growth of CdTe in conventional MBE mode, and ZnTe in atomic-layer epitaxy (ALE) mode. The latter is achieved by depositing Te and Zn subsequently instead of simultaneously, which results in the self-limiting formation of a half-monolayer Zn on a monolayer Te \cite{takemura1992}. The half-filled layer of Zn atoms is completed with Cd atoms during the subsequent CdTe MBE step, resulting in one monolayer $\text{Cd}_{0.5}\text{Zn}_{0.5}\text{Te}$ embedded in the CdTe.  The layer sequence of a period of a SLS is shown in Fig.~\ref{fig:1}(b). A HRXRD $2 \theta - \omega$ scan of the (004) reflection of a SLS, and the simulated intensity profile are shown in Fig.~\ref{fig:1}(a). Narrow peaks and numerous satellites (labeled $\pm1,\ \pm2,\ \pm3$...) indicate high crystal quality, uniform superlattice periods and abrupt interfaces, despite the fact that growth is performed on highly lattice-mismatched GaAs substrate material. The SLS period $p$, and consequently its total thickness (typically in the range of $1-\SI{3}{\micro\metre}$) are inferred from the angular spacings of the SLS satellites. By balancing of the forces acting within a single superlattice period and taking into account the self-limiting nature of the ALE growth process \cite{dunstan1997}\footnote{See supplementary online material for a brief derivation of Eq.~1.}, one readily derives
\begin{equation}
 a_\text{SLS} = a_\text{CT}\ \left[1 + \frac{f}{1 + \alpha \left(t_\text{CT} \times \varphi_\text{Te}\right)}\right],
\label{eq:1}
\end{equation}
which relates the effective lattice constant $a_\text{SLS}$ of a SLS to the product of CdTe-layer growth time, $t_\text{CT}$, and Te beam-equivalent pressure, $\varphi_\text{Te}$, which are both straightforwardly accessible in the experiment. In Eq.~\ref{eq:1}, $f$ is the lattice mismatch between unstrained CdTe and $\text{Cd}_{0.5}\text{Zn}_{0.5}\text{Te}$ and $a_\text{CT}$ is the lattice constant of CdTe \cite{Schenk1996}. The parameter $\alpha$ contains material (stiffnesses $c_{ij}$ \cite{Andrusiv1983} and $\text{Cd}_{0.5}\text{Zn}_{0.5}\text{Te}$ epilayer thickness), and process-specific parameters (normalized CdTe growth speed). The effective lattice constant $a_\text{SLS}$ is deduced from the angular spacing of the GaAs substrate (``S'') and the zero-order Bragg reflection of the SLS (``0''). Fig.~\ref{fig:1}(c) shows the obtained $a_\text{SLS}$ for a set of SLS as a function of $t_\text{CT} \times \varphi_\text{Te}$. A fit of Eq.~\ref{eq:1}, with $\alpha = \SI{3.4e5}{\per\second\per\torr}$ (black line) is in good agreement with the data. Thus, $a_\text{SLS}$ can be controlled over a wide range by simply adjusting $t_\text{CT} \times \varphi_\text{Te}$. This degree of freedom allows for a precise control of the strain in HgTe (001) epilayers and, in turn, offers new ways to modify the band structure of bulk layers and QWs.\par

\begin{figure}[b]
\includegraphics{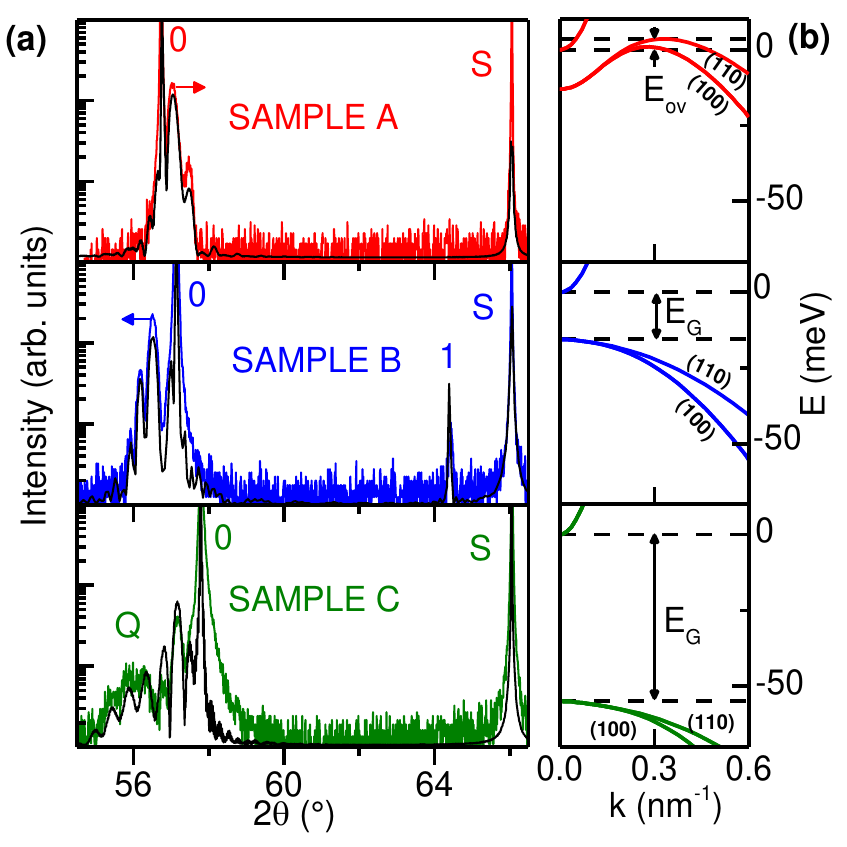}
\caption{\label{fig:2} (a) HRXRD $2 \theta - \omega$ scan of the (004) reflection of sample A (top), B (center) and C (bottom). Reflections are labeled as in Fig.~\ref{fig:1}(a). Arrows indicate the strain-induced shift of the $\text{Cd}_{0.7}\text{Hg}_{0.3}\text{Te}$ QW barrier reflections (unlabeled). The diffracted intensity of the QW (``Q'') is only visible in sample C. Black lines are simulated diffraction profiles which are slightly offset downwards for clarity. (b) Calculated band structures of samples A, B and C (from top to bottom). Numbers in brackets denote crystal directions. Dashed lines indicate the energetic overlap $E_\text{ov}$ of VB and CB for sample A, and the band gap $E_\text{G}$ of samples B and C.}
\end{figure}

To demonstrate the scope of the modifications of the band structure, we have fabricated a set of three QWs A, B and C, with distinct strain and thickness parameters for magnetotransport measurements. Samples A and B are thick QWs with almost identical thickness ($d_\text{QW} = \SI{16}{\nano\metre}$ and $\SI{15}{\nano\metre}$), and similar top- and bottom barrier layers ($\text{Cd}_{0.7}\text{Hg}_{0.3}\text{Te}$ with $d_\text{barr} = \SI{17}{\nano\metre}$, each). The virtual substrates, however, are different. Sample A is grown on a thick, relaxed CdTe (001) epilayer grown on GaAs, which gives rise to tensile strain. Samples B and C are grown on two different SLS, that induce moderate and large compressive strain on the respective QWs. In sample C, a solid solution of (Zn,Cd,Hg)Te is used as barrier material ($d_\text{barr} = \SI{18}{\nano\metre}$) (instead of the standard $\text{Cd}_{0.7}\text{Hg}_{0.3}\text{Te}$), to lower the mismatch between substrate and barriers, thus avoiding relaxation of the heterostructure. The QW thickness of sample C is $d_\text{QW} = \SI{7.5}{\nano\metre}$.  HRXRD $2\theta - \omega$ scans of the (004) diffraction profiles of all three samples are shown in Fig.~\ref{fig:2}(a). The color-coding of data (red: sample A, blue: sample B green: sample C) holds for the rest of this work. The strain in the HgTe layers, deduced from the S - 0 angular separation, is $\varepsilon =  -\SI{0.3}{\percent}$ for sample A, $+\SI{0.4}{\percent}$ for sample B, and $+\SI{1.4}{\percent}$ for sample C. Unlabeled reflections are caused by the (Zn,Cd,Hg)Te barriers. The barriers and the QW of all samples are fully strained, as verified by comparing the diffraction profiles with appropriate simulations (black lines). It is worth noting, that the symmetric measurement geometry probes the out-of-plane response of the lattice constants to the in-plane strain. The magnitude of the response is determined by the lattice constant mismatch of SLS-barrier and SLS-QW, respectively, and the Poisson's ratios of the materials. Arrows highlight the strain-induced shift of the barrier reflection of samples A and B. Note that relaxation of the HgTe layer would be seen as a lowered shift of the topbarrier reflection \footnote{See supplementary online material for additional remarks on the determination of the state of strain in the QW from HRXRD measurements and fits.}. From its angular position, the composition of the barriers of sample C can be estimated as $\text{Zn}_{0.20}\text{Cd}_{0.56}\text{Hg}_{0.24}\text{Te}$. Remarkably, due to the large mismatch between QW and barriers, the QW of sample C is directly visible in the diffraction pattern as an isolated set of fringes [labelled ``Q'' in Fig.~\ref{fig:2}(a), bottom]. Fig.~\ref{fig:2}(b) shows the band structures of the three QWs, calculated using an eight-band $\mathbf{k \cdot p}$ model \cite{Novik2005}. The variety in energy dispersions accessible by varying the strain and the thickness of the QW is evident. Upon comparing samples A and B, one observes that the strain in the layers primarily affects the shape of the valence band (VB) and causes a transition from a semimetal-like system with energetic overlap $E_\text{ov}$ between VB and conduction band (CB) to a direct-band-gap semiconductor. From the band structure of samples B and C, one sees that the total band gap $E_\text{G}$ increases significantly when the compressive strain is increased. Since the QW thickness of all samples is well above $d_\text{c}$, the band ordering is topologically nontrivial. Thus, sample A is a topological two-dimensional semimetal, and samples B and C are expected to be 2D-TIs.\par

\begin{figure}[b]
\includegraphics{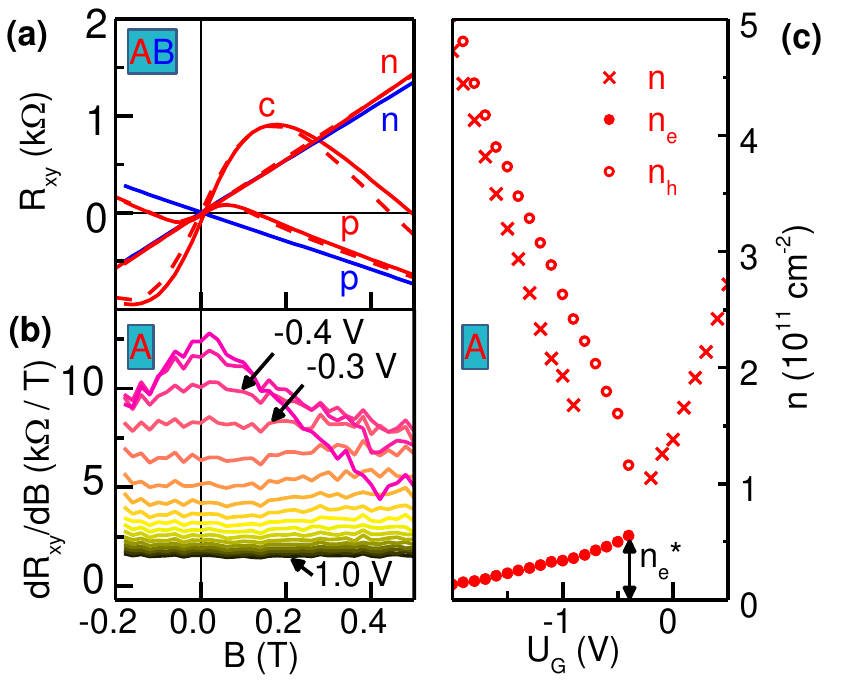}
\caption{\label{fig:3} (a) Hall resistance $R_\text{xy}$ as function of magnetic field for sample A (red) and B (blue) at carrier densities: $n \approx +2$ and $-\SI{4e11}{\per\centi\metre\squared}$, labelled n and p, respectively. Measurement at  $-\SI{2e11}{\per\centi\metre\squared}$ (labelled c) was only possible for sample A. Dashed lines are obtained by fitting $R_\text{xy}$ and $R_\text{xx}$ simultaneously to a two-carrier Drude model. (b) Derivative of the Hall resistance of sample A as function of magnetic field B for gate voltages from -0.6 V (magenta) to +1.0 V (black). (c) Crosses: net carrier density estimated from $R_\text{xy}$ at higher fields, as a function of gate voltage $U_\text{G}$. Filled (empty) circles: density of electrons (holes) obtained from two-carrier fit. The double arrow highlights n-type carrier density $n_\text{e}^*$ at the onset of two-carrier conductance.}
\end{figure}

These characteristic band dispersion properties are reflected qualitatively and quantitatively in distinct magneto-transport features. Measurements were carried out on top-gated Hall bar devices fabricated using optical lithography. In a first set of experiments (Fig.~\ref{fig:3}), we compare the behavior of samples A and B. Varying the gate voltage $U_\text{G}$ from negative to positive values shifts the Fermi energy from the VB into the CB. This is apparent by a transition from p-conducting to n-conducting behavior that is reflected in a sign-change of the Hall resistance $R_\text{xy}$. As shown in Fig.~\ref{fig:3}(a),  the presence or absence of a band overlap in samples A and B results in markedly different characteristics of $R_\text{xy}$ (traces with similar labels are chosen such that the carrier densities are equal within experimental resolution). When the Fermi level is deep in the CB, the Hall resistance in both samples is purely electron-like, and both traces are linear (traces labelled ``n''). As the gate voltage is lowered, a pronounced curvature is observed in $R_\text{xy}$ for sample A (trace ``c''). This is characteristic of a system with coexisting electron- and hole-like carriers of different mobilities \cite{Ashcroft1976}, and indicates an overlap of the CB and VB \cite{kvon2008}. In the same gate voltage regime, sample B is highly resistive, and no Hall voltage measurement is possible, implying  that the Fermi energy is in the band gap. Finally, for strong negative gate voltages, an entirely linear trace is recovered for sample B, while, in contrast, two-carrier conductance persists in sample A (traces ``p''). We interpret this as reflecting the effective pinning of the Fermi level at the van Hove singularity (``camel's back'') in the VB density of states in sample A.\par

Our data allows for a more detailed analysis of the evolving electron- and hole densities of sample A. As soon as the Fermi energy intersects with the VB, two-carrier conductance sets in, and non constant $dR_\text{xy} / dB$ is evidence that, at low fields, the Hall resistance is no longer described by the simple single-carrier expression $R_\text{xy} = B\ (ne)^{-1}$. Experimentally, we observe this effect for negative gate voltages larger than $U_\text{G} = -\SI{0.4}{\volt}$ [Fig.~\ref{fig:3}(b)]. The onset of two-carrier conductance allows us to estimate the energy overlap $E_\text{ov}$ of the VB and CB. A simultaneous fit of $R_\text{xy}$ and $R_\text{xx}$ to the standard two-carrier Drude model \cite{Ashcroft1976} yields the density of electron- and hole-like carriers $n_\text{e}$ and $n_\text{h}$. Fits are shown in Fig.~\ref{fig:3}(a) as dashed lines. The resulting densities for the whole measurement set are shown in Fig.~\ref{fig:3}(c), together with the net density $n$, extracted from $R_\text{xy}$ at higher fields. The n-type carrier density at the onset of two carrier conductance (black arrow) is $n_e^* = \SI{5.5e11}{\per\centi\metre\squared}$. Using $E_\text{ov} = E_\text{F} = n_e^*\ \pi \hbar^2 m_\text{e}^{-1}$, we obtain $E_\text{ov} = \SI{4.5}{\milli\electronvolt}$ for the band overlap, which is slightly larger than the value inferred from band structure calculations [$E_\text{ov} = \SI{3.5}{\milli\electronvolt}$, see Fig.~\ref{fig:2}(c), top]. The electron effective mass in the CB is taken as $m_\text{e} = 0.028\ m_0$ ($m_0$ is the mass of a free electron), in agreement with the $\mathbf{k \cdot p}$ model calculations.\par  

\begin{figure}[b]
\includegraphics{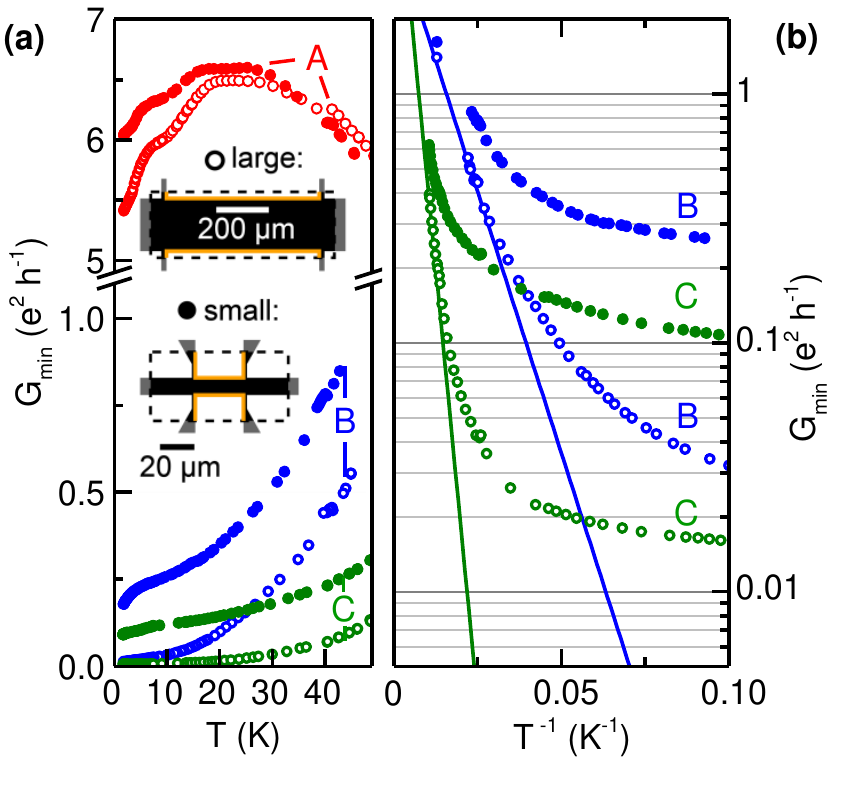}
\caption{\label{fig:4} (a) Minimum conductance $G_\text{min}$ of samples A, B and C as function of temperature. Empty and filled dots correspond to values measured on large and small Hall bars. Inset gives the Hall bar dimensions - note the different scales. Dashed rectangles and black areas indicate the gated regions, while the probed edge current paths are indicated by orange lines. (b) Arrhenius plot of $G_\text{min}$ of samples B and C at $T > \SI{10}{\kelvin}$. Lines are fit of the large Hall bar data to Eq.~\ref{eq:2}.}
\end{figure}

Finally, we discuss thermal activation studies of conductance, which allows us to discriminate between metallic sheet conductance at the charge neutrality point (sample A) and edge state conductance (samples B and C), and to estimate the magnitude of the strain-induced band gap of samples B and C. To distinguish between current flowing in the bulk of the QW and one-dimensional edge current, two Hall bars with different dimensions have been fabricated for each QW [inset in Fig.~\ref{fig:4}(a)]. Whereas the width-to-length ratio is identical ($w\ /\ l = 1\ /\ 3$) for both Hall bars, the length of the gated edge changes by roughly a factor of ten ($l_\text{edge} = 58\ \upmu\text{m}$ and $620\ \upmu\text{m}$). For temperatures in the range from $T = \SI{1.8}{\kelvin}$ to $\SI{90}{\kelvin}$, minimum values of the longitudinal conductance $G_\text{min}$ are measured at gate voltages corresponding to the situation when the Fermi energy is located at the charge neutrality point for sample A and the mid band gap position for samples B and C. The results are plotted in Fig.~\ref{fig:4}(a). Sample A (red curves) shows a high $G_\text{min}$, which changes only moderately with temperature. The observed low-temperature increase of $G_\text{min}$ with temperature was reported previously in Ref.~\cite{olshanetsky2014} and has been attributed to long-range disorder scattering \cite{Knap2014}. We suggest that the decrease in $G_\text{min}$ at higher $T$ is due to enhanced phonon-scattering. The fact that $G_\text{min}$ is qualitatively similar for the large and small Hall bar indicates that the current flows in the bulk of the QW (as mentioned above, $w\ /\ l$, relating two-dimensional conductance to conductivity, is the same in both devices). The behavior of samples B and C is significantly different. For all temperatures, $G_\text{min}$ is much smaller than in sample A, and a thermally activated increase in conductance is observed, as typical for semiconductors. A logarithmic plot of the high temperature data ($T > \SI{10}{\kelvin}$) of samples B and C versus $T^{-1}$ is shown in Fig.~\ref{fig:4}(b). As a clear indication of edge channel transport in the low temperature regime, we observe that $G_\text{min}$ of both samples tends to saturate, and the saturation values of large and small Hall bar roughly scale with the inverse of the edge channel length (10 / 1). Since the edge length of both Hall bars significantly exceeds the inelastic mean free path of the QSH edge channels \cite{konig2007}, and thus the number of scattering events is approximately proportional to the length of the channel \cite{vayrynen2013}, this is an expected signature of edge channel transport. With increasing temperature, the thermally activated conductance over the whole area of the mesa becomes dominant. It is possible to extract the band gap from the conductance in the high temperature regime. By fitting the measured $G_\text{min}$ of the large Hall bar to
\begin{equation}
G_\text{min} = G_0\ \exp \left(-\frac{E_\text{G}}{2 k_\text{B} T}\right),
\label{eq:2}
\end{equation}
we obtain $E_\text{G} = \SI{17}{\milli\electronvolt}$ and $\SI{55}{\milli\electronvolt}$ for samples B and C [solid lines in Fig.~\ref{fig:4}(b)], in good agreement with band structure calculations [$E_\text{G} = \SI{16}{\milli\electronvolt}$ and $\SI{55}{\milli\electronvolt}$, see Fig.~\ref{fig:2}(b), center and bottom]. Reliable fits of Eq.~\ref{eq:2} are only possible for the large Hall bars, since the QSH edge state conductance of the small Hall bars substantially contributes to the total conductance even at high temperatures. \par

In conclusion, we have presented a method to significantly increase the band gap of HgTe based 2D-TIs, based on strain-engineering via dedicated $\text{CdTe-Cd}_{0.5}\text{Zn}_{0.5}\text{Te}$ SLS virtual substrates. In particular, we have shown that applying compressive strain to QWs results in energy gaps as high as $\SI{55}{\milli\electronvolt}$. This value is the largest ever reported in inverted ($d_\text{QW} > \SI{6.3}{\nano\metre}$) HgTe QWs, is well above $k_\text{B}T$ at room temperature, and is a necessary step towards room temperature QSH-based electronic devices. Furthermore, we have demonstrated that thick QWs can be transformed from semimetals to 2D-TIs by changing their strain from tensile to compressive. Finally, we emphasize the accuracy of strain control via the SLS approach. The effective lattice constant of the SLS can be conveniently controlled by the product of CdTe-period growth time and Te beam-equivalent pressure, with both parameters straightforwardly accessible in crystal growth.

\begin{acknowledgments}
This work was supported by the German Research Foundation DFG (SPP 1666, SFB 1170), the U.S. DARPA Meso project (grant N66001-11-1-4105), the European Research Council (advanced grant project 3-TOP), the Helmholtz Foundation (VITI) and the Elitenetzwerk Bayern (ENB)
\end{acknowledgments}

%

\pagebreak
\widetext
\begin{center}
\textbf{\large Strain Engineering of the Band Gap of HgTe Quantum Wells using Superlattice Virtual Substrates\\ Supplementary Material}
\end{center}

\setcounter{equation}{0}
\setcounter{figure}{0}
\setcounter{table}{0}
\setcounter{page}{1}
\makeatletter
\renewcommand{\theequation}{S\arabic{equation}}
\renewcommand{\thefigure}{S\arabic{figure}}
\renewcommand{\bibnumfmt}[1]{[S#1]}
\renewcommand{\citenumfont}[1]{S#1}

\subsection{Introduction}
In this supplementary section, we show how to derive Eq.~1 of the main article from basic elasticity theory, and by considering the growth mechanisms of the experiment. Furthermore, we demonstrate that the (004) High-resolution-X-ray-diffraction (HRXRD) $2 \theta-\omega$ measurements and fits, shown in Fig.~2 of the main article, provide information of the state of strain of the HgTe layer in all three spatial directions, even though the layer itself is not directly visible for samples A and B. 

\subsection{Derivation of equation 1}
As already noted in the main text, Eq.~1 can be obtained by balancing of the forces of a single bilayer of the strained-layer-superlattice (SLS). Consider a SLS consisting of two cubic materials, with lattice constants $a_1$, $a_2$, and elastic constants $c_{ij,1}$, $c_{ij,2}$. If the individual layer thicknesses $d_1$ and $d_2$ are below the critical thickness of the particular material system, the growth will take place coherently, i.e. the layers will be strained, such that the in-plane lattice constant $a_\text{SLS}$ is similar. For the (001) growth direction, the amount of stress acting parallel to the bilayer interface can be calculated using Eq.~11 from Ref. \cite{dunstan1997} as 
\begin{equation}
 \sigma = M \varepsilon_{||}.
\label{eq:1}
\end{equation}
Here, the stresses in both in-plane directions are taken as similar, and the elastic constants are absorbed in $M = c_{11} + c_{12} - 2\ c_{12}^2/c_{11}$. The level of strain in each layer depends on the material type and the in-plane lattice constant as $\varepsilon_{||1,2} = (a_{1,2}-a_\text{SLS})\ a_{1,2}^{-1}$. In equilibrium, the forces $f_{1,2} = \sigma_{1,2}\ d_{1,2}$ in the two layers have to cancel each other. Using this condition, one finds
\begin{equation}
 a_\text{SLS} = a_1 \left(1 + \frac{f}{1+ \frac{M_1}{M_2}\frac{d_1}{d_2}}\right)
\label{eq:2}
\end{equation}
for the lattice constant in equilibrium. Here, $f = (a_2-a_1)a_1^{-1}$ is the lattice mismatch between the two materials. In order to apply Eq.~\ref{eq:2} to our growth process, we set material one as CdTe and material 2 as $\text{Cd}_{0.5}\text{Zn}_{0.5}\text{Te}$ and define all material constants accordingly, using Refs. \cite{Schenk1996,Andrusiv1983}. Due to the self-limiting ALE process, $d_2$ is fixed. Since we grow the CdTe layer in Cd-rich conditions, its thickness $d_1$ is proportional to the Te-flux $\phi_\text{Te}$ and the CdTe growth time $t_\text{CT}$, i.e. $d_1 = v\ (t_\text{CT} \times \phi_\text{Te})$. Here, $v$ is the (unknown) growth speed of the CdTe layer. Substituting this in Eq.~\ref{eq:2} and absorbing all material constants and $v$ into the fitting parameter $\alpha$ yields Eq.~1 of the main article. The equation holds for SLS, which consist of multiples of bilayers, if the SLS can be considered as ``free standing'', i.e. if the substrate does not exert considerable strain in the SLS. This is the case in our samples, since the SLS relax within the first few monolayers of growth, due to the large mismatch between the GaAs substrate and the SLS. We have confirmed this by HRXRD reciprocal space maps of the (115) reflection of our structures. \par

\subsection{The state of strain of quantum wells from X-ray measurements}

\begin{figure}[bt]
\includegraphics{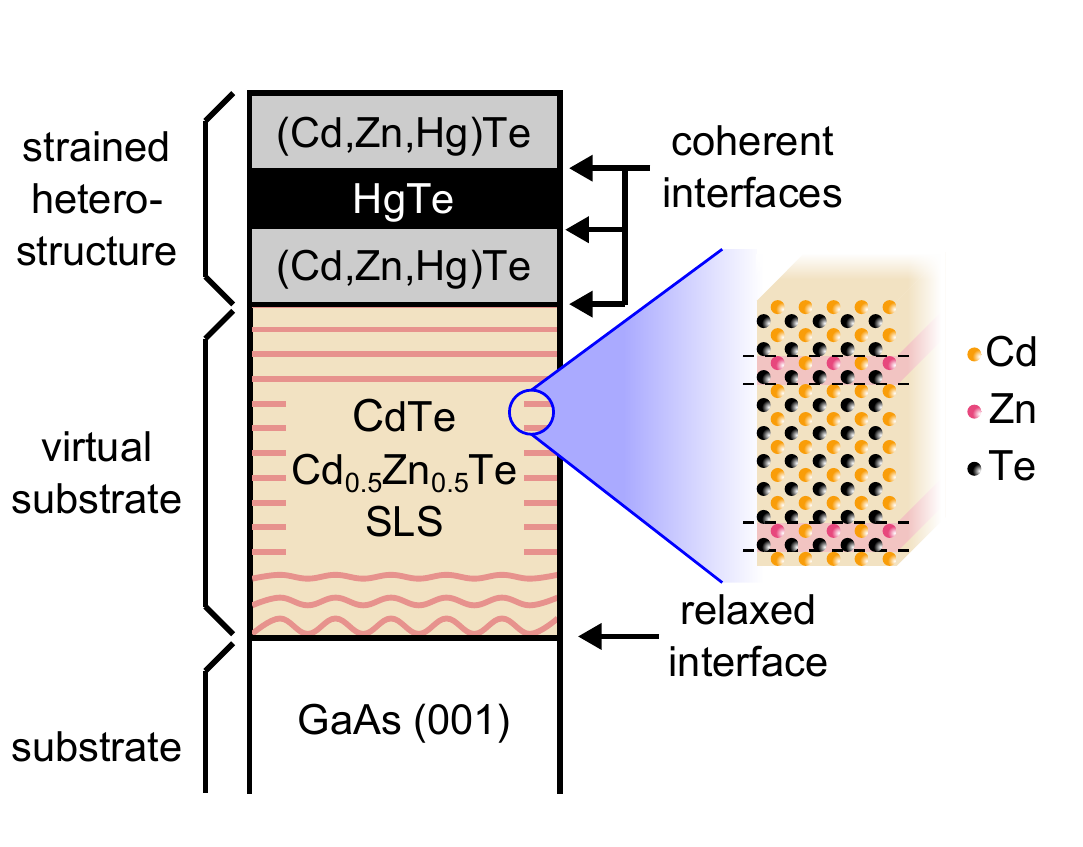}
\caption{\label{fig:s1} Illustration of the sample layout of the three heterostructures discussed in the main text. Closeup shows single period of SLS. Note that the virtual substrate of sample A is a CdTe single layer.}
\end{figure}

We now turn to a more detailed discussion of the HRXRD analysis of the state of strain in the HgTe quantum wells (QWs). For clarity, the sample layout is depicted in Fig.~\ref{fig:s1}. Regarding samples A and B, the HgTe layers are not directly seen in the HRXRD measurements (Fig.~2a), since the intensity of their reflections is overwhelmed by the intensity of the $\text{Cd}_{0.7}\text{Hg}_{0.3}\text{Te}$ barriers. Nonetheless, their state of strain can be inferred without a doubt for all spatial directions. Since the (004) reflection which we are analyzing is symmetric (i.e. $\omega = 2\ \theta$ in the experimental setup), the measurements shown in Fig.~2a of the main article only probe the out-of-plane lattice constants. This lattice constant is determined by the actual material composition and the out-of-plane strain, which arises due to the crystal's response to the in-plane strain, and can be determined from Poisson's ratio. The simulated intensity profiles shown in Fig.~2a are obtained assuming fully coherent growth (``coherent interfaces'' in Fig.~\ref{fig:s1}), which means that the barriers and the QW adopt the in-plane lattice constant of the SLS. The agreement between simulation and measurement confirms this assumption.\par

\begin{figure}[bt]
\includegraphics{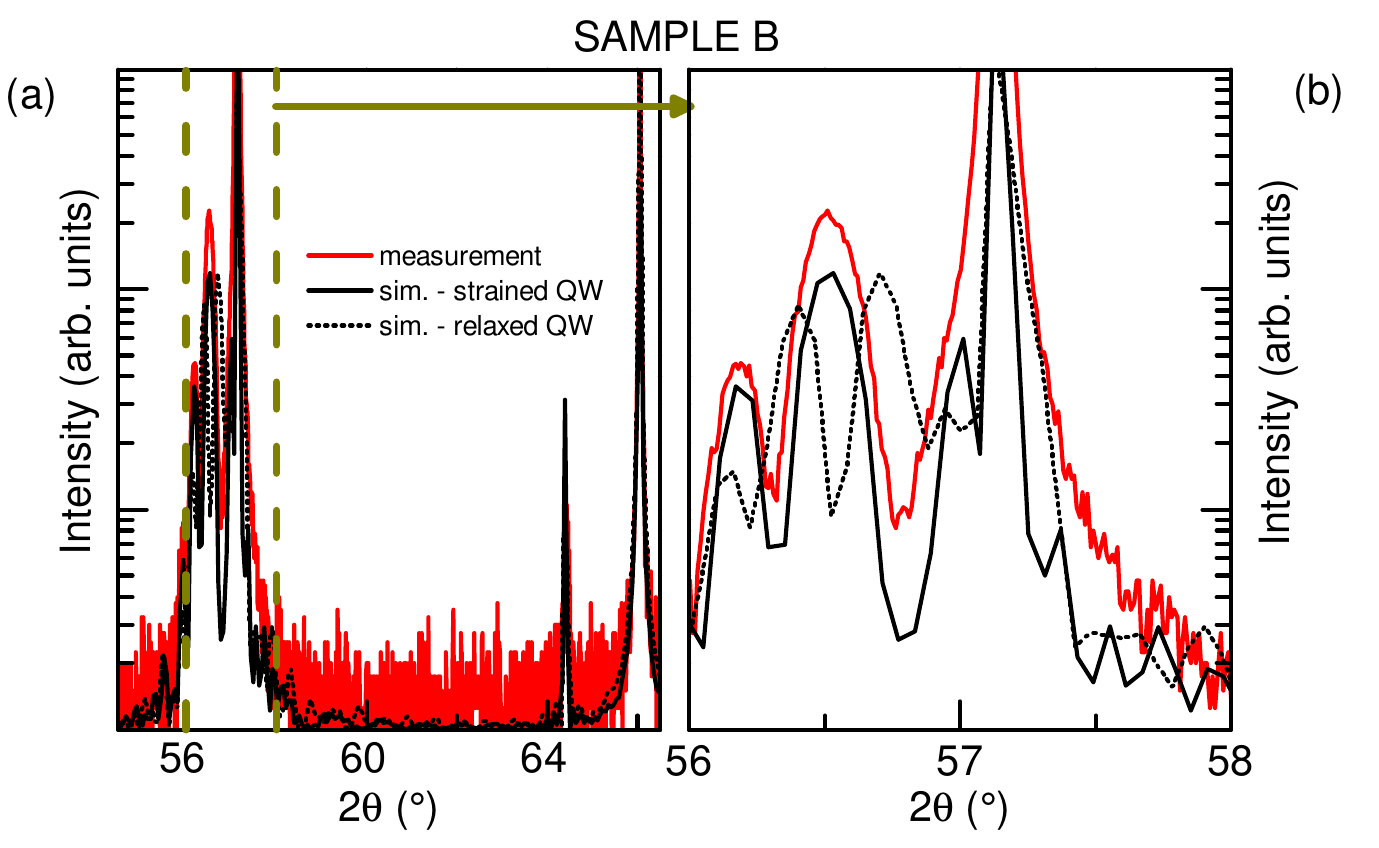}
\caption{\label{fig:s2}HRXRD $2 \theta - \omega$ scan of the (004) reflection of sample B. Black lines are simulated diffraction profiles which are slightly offset downwards for clarity, and are obtained assuming a fully strained (solid) and fully relaxed (dotted) HgTe QW. Scales in (a) are similar to Fig.~2a of the main manuscript. (b) shows closeup of the reflections of SLS and (Cd,Hg)Te layers.}
\end{figure}

Furthermore, any relaxation in the HgTe layer would cause the topbarrier to adopt to the altered in-plane lattice constant of this partially relaxed layer. This would cause a change in the strain-induced shift of the fringe of the topbarrier. As demonstrated for sample B in Fig.~\ref{fig:s2}, this clearly does not occur. Thus it is safe to infer that the HgTe layer fully adopts the lattice constant of the SLS in-plane and, similar to the barrier material, its out-of-plane lattice constant is modified according to Poisson's ratio. Additionally, we again point out that we directly see the HgTe reflection in sample C, where the out-of-plane state of strain of the HgTe can be directly confirmed by the position of its fringes (labelled ``Q'' in bottom Fig.~2a of the main article). Since the induced strain is largest in this sample, we are even more confident that the HgTe is also fully strained in samples A and B.


\begin{thebibliography}{21}%
\makeatletter
\providecommand \@ifxundefined [1]{%
 \@ifx{#1\undefined}
}%
\providecommand \@ifnum [1]{%
 \ifnum #1\expandafter \@firstoftwo
 \else \expandafter \@secondoftwo
 \fi
}%
\providecommand \@ifx [1]{%
 \ifx #1\expandafter \@firstoftwo
 \else \expandafter \@secondoftwo
 \fi
}%
\providecommand \natexlab [1]{#1}%
\providecommand \enquote  [1]{``#1''}%
\providecommand \bibnamefont  [1]{#1}%
\providecommand \bibfnamefont [1]{#1}%
\providecommand \citenamefont [1]{#1}%
\providecommand \href@noop [0]{\@secondoftwo}%
\providecommand \href [0]{\begingroup \@sanitize@url \@href}%
\providecommand \@href[1]{\@@startlink{#1}\@@href}%
\providecommand \@@href[1]{\endgroup#1\@@endlink}%
\providecommand \@sanitize@url [0]{\catcode `\\12\catcode `\$12\catcode
  `\&12\catcode `\#12\catcode `\^12\catcode `\_12\catcode `\%12\relax}%
\providecommand \@@startlink[1]{}%
\providecommand \@@endlink[0]{}%
\providecommand \url  [0]{\begingroup\@sanitize@url \@url }%
\providecommand \@url [1]{\endgroup\@href {#1}{\urlprefix }}%
\providecommand \urlprefix  [0]{URL }%
\providecommand \Eprint [0]{\href }%
\providecommand \doibase [0]{http://dx.doi.org/}%
\providecommand \selectlanguage [0]{\@gobble}%
\providecommand \bibinfo  [0]{\@secondoftwo}%
\providecommand \bibfield  [0]{\@secondoftwo}%
\providecommand \translation [1]{[#1]}%
\providecommand \BibitemOpen [0]{}%
\providecommand \bibitemStop [0]{}%
\providecommand \bibitemNoStop [0]{.\EOS\space}%
\providecommand \EOS [0]{\spacefactor3000\relax}%
\providecommand \BibitemShut  [1]{\csname bibitem#1\endcsname}%
\let\auto@bib@innerbib\@empty
\bibitem [{\citenamefont {K{\"o}nig}\ \emph {et~al.}(2007)\citenamefont
  {K{\"o}nig}, \citenamefont {Wiedmann}, \citenamefont {Br{\"u}ne},
  \citenamefont {Roth}, \citenamefont {Buhmann}, \citenamefont {Molenkamp},
  \citenamefont {Qi},\ and\ \citenamefont {Zhang}}]{konig2007}%
  \BibitemOpen
  \bibfield  {author} {\bibinfo {author} {\bibfnamefont {M.}~\bibnamefont
  {K{\"o}nig}}, \bibinfo {author} {\bibfnamefont {S.}~\bibnamefont {Wiedmann}},
  \bibinfo {author} {\bibfnamefont {C.}~\bibnamefont {Br{\"u}ne}}, \bibinfo
  {author} {\bibfnamefont {A.}~\bibnamefont {Roth}}, \bibinfo {author}
  {\bibfnamefont {H.}~\bibnamefont {Buhmann}}, \bibinfo {author} {\bibfnamefont
  {L.~W.}\ \bibnamefont {Molenkamp}}, \bibinfo {author} {\bibfnamefont {X.-L.}\
  \bibnamefont {Qi}}, \ and\ \bibinfo {author} {\bibfnamefont {S.-C.}\
  \bibnamefont {Zhang}},\ }\href@noop {} {\bibfield  {journal} {\bibinfo
  {journal} {Science}\ }\textbf {\bibinfo {volume} {318}},\ \bibinfo {pages}
  {766} (\bibinfo {year} {2007})}\BibitemShut {NoStop}%
\bibitem [{\citenamefont {Roth}\ \emph {et~al.}(2009)\citenamefont {Roth},
  \citenamefont {Br{\"u}ne}, \citenamefont {Buhmann}, \citenamefont
  {Molenkamp}, \citenamefont {Maciejko}, \citenamefont {Qi},\ and\
  \citenamefont {Zhang}}]{roth2009}%
  \BibitemOpen
  \bibfield  {author} {\bibinfo {author} {\bibfnamefont {A.}~\bibnamefont
  {Roth}}, \bibinfo {author} {\bibfnamefont {C.}~\bibnamefont {Br{\"u}ne}},
  \bibinfo {author} {\bibfnamefont {H.}~\bibnamefont {Buhmann}}, \bibinfo
  {author} {\bibfnamefont {L.~W.}\ \bibnamefont {Molenkamp}}, \bibinfo {author}
  {\bibfnamefont {J.}~\bibnamefont {Maciejko}}, \bibinfo {author}
  {\bibfnamefont {X.-L.}\ \bibnamefont {Qi}}, \ and\ \bibinfo {author}
  {\bibfnamefont {S.-C.}\ \bibnamefont {Zhang}},\ }\href@noop {} {\bibfield
  {journal} {\bibinfo  {journal} {Science}\ }\textbf {\bibinfo {volume}
  {325}},\ \bibinfo {pages} {294} (\bibinfo {year} {2009})}\BibitemShut
  {NoStop}%
\bibitem [{\citenamefont {Br{\"u}ne}\ \emph {et~al.}(2012)\citenamefont
  {Br{\"u}ne}, \citenamefont {Roth}, \citenamefont {Buhmann}, \citenamefont
  {Hankiewicz}, \citenamefont {Molenkamp}, \citenamefont {Maciejko},
  \citenamefont {Qi},\ and\ \citenamefont {Zhang}}]{brune2012}%
  \BibitemOpen
  \bibfield  {author} {\bibinfo {author} {\bibfnamefont {C.}~\bibnamefont
  {Br{\"u}ne}}, \bibinfo {author} {\bibfnamefont {A.}~\bibnamefont {Roth}},
  \bibinfo {author} {\bibfnamefont {H.}~\bibnamefont {Buhmann}}, \bibinfo
  {author} {\bibfnamefont {E.~M.}\ \bibnamefont {Hankiewicz}}, \bibinfo
  {author} {\bibfnamefont {L.~W.}\ \bibnamefont {Molenkamp}}, \bibinfo {author}
  {\bibfnamefont {J.}~\bibnamefont {Maciejko}}, \bibinfo {author}
  {\bibfnamefont {X.-L.}\ \bibnamefont {Qi}}, \ and\ \bibinfo {author}
  {\bibfnamefont {S.-C.}\ \bibnamefont {Zhang}},\ }\href@noop {} {\bibfield
  {journal} {\bibinfo  {journal} {Nature Physics}\ }\textbf {\bibinfo {volume}
  {8}},\ \bibinfo {pages} {485} (\bibinfo {year} {2012})}\BibitemShut {NoStop}%
\bibitem [{\citenamefont {Br{\"u}ne}\ \emph {et~al.}(2011)\citenamefont
  {Br{\"u}ne}, \citenamefont {Liu}, \citenamefont {Novik}, \citenamefont
  {Hankiewicz}, \citenamefont {Buhmann}, \citenamefont {Chen}, \citenamefont
  {Qi}, \citenamefont {Shen}, \citenamefont {Zhang},\ and\ \citenamefont
  {Molenkamp}}]{brune2011}%
  \BibitemOpen
  \bibfield  {author} {\bibinfo {author} {\bibfnamefont {C.}~\bibnamefont
  {Br{\"u}ne}}, \bibinfo {author} {\bibfnamefont {C.~X.}\ \bibnamefont {Liu}},
  \bibinfo {author} {\bibfnamefont {E.~G.}\ \bibnamefont {Novik}}, \bibinfo
  {author} {\bibfnamefont {E.~M.}\ \bibnamefont {Hankiewicz}}, \bibinfo
  {author} {\bibfnamefont {H.}~\bibnamefont {Buhmann}}, \bibinfo {author}
  {\bibfnamefont {Y.~L.}\ \bibnamefont {Chen}}, \bibinfo {author}
  {\bibfnamefont {X.~L.}\ \bibnamefont {Qi}}, \bibinfo {author} {\bibfnamefont
  {Z.~X.}\ \bibnamefont {Shen}}, \bibinfo {author} {\bibfnamefont {S.~C.}\
  \bibnamefont {Zhang}}, \ and\ \bibinfo {author} {\bibfnamefont {L.~W.}\
  \bibnamefont {Molenkamp}},\ }\href@noop {} {\bibfield  {journal} {\bibinfo
  {journal} {Physical Review Letters}\ }\textbf {\bibinfo {volume} {106}},\
  \bibinfo {pages} {126803} (\bibinfo {year} {2011})}\BibitemShut {NoStop}%
\bibitem [{\citenamefont {Br{\"u}ne}\ \emph {et~al.}(2014)\citenamefont
  {Br{\"u}ne}, \citenamefont {Thienel}, \citenamefont {Stuiber}, \citenamefont
  {B{\"o}ttcher}, \citenamefont {Buhmann}, \citenamefont {Novik}, \citenamefont
  {Liu}, \citenamefont {Hankiewicz},\ and\ \citenamefont
  {Molenkamp}}]{brune2014}%
  \BibitemOpen
  \bibfield  {author} {\bibinfo {author} {\bibfnamefont {C.}~\bibnamefont
  {Br{\"u}ne}}, \bibinfo {author} {\bibfnamefont {C.}~\bibnamefont {Thienel}},
  \bibinfo {author} {\bibfnamefont {M.}~\bibnamefont {Stuiber}}, \bibinfo
  {author} {\bibfnamefont {J.}~\bibnamefont {B{\"o}ttcher}}, \bibinfo {author}
  {\bibfnamefont {H.}~\bibnamefont {Buhmann}}, \bibinfo {author} {\bibfnamefont
  {E.~G.}\ \bibnamefont {Novik}}, \bibinfo {author} {\bibfnamefont {C.-X.}\
  \bibnamefont {Liu}}, \bibinfo {author} {\bibfnamefont {E.~M.}\ \bibnamefont
  {Hankiewicz}}, \ and\ \bibinfo {author} {\bibfnamefont {L.~W.}\ \bibnamefont
  {Molenkamp}},\ }\href@noop {} {\bibfield  {journal} {\bibinfo  {journal}
  {Physical Review X}\ }\textbf {\bibinfo {volume} {4}},\ \bibinfo {pages}
  {041045} (\bibinfo {year} {2014})}\BibitemShut {NoStop}%
\bibitem [{Note1()}]{Note1}%
  \BibitemOpen
  \bibinfo {note} {Alternatively, several $\upmu \protect \text {m}$ thick
  buffer layers of fully relaxed CdTe on Si or GaAs can be
  employed}\BibitemShut {NoStop}%
\bibitem [{\citenamefont {Pfeuffer-Jeschke}(2000)}]{PfeufferJeschke2000}%
  \BibitemOpen
  \bibfield  {author} {\bibinfo {author} {\bibfnamefont {A.}~\bibnamefont
  {Pfeuffer-Jeschke}},\ }\emph {\bibinfo {title} {{Bandstruktur und
  Landau-Niveaus quecksilberhaltiger II-VI-Heterostrukturen}}},\ \href@noop {}
  {Ph.D. thesis},\ \bibinfo  {school} {Universit{\"a}t W{\"u}rzburg} (\bibinfo
  {year} {2000})\BibitemShut {NoStop}%
\bibitem [{\citenamefont {Feldman}\ \emph {et~al.}(1986)\citenamefont
  {Feldman}, \citenamefont {Austin}, \citenamefont {Dayem},\ and\ \citenamefont
  {Westerwick}}]{feldman1986}%
  \BibitemOpen
  \bibfield  {author} {\bibinfo {author} {\bibfnamefont {R.}~\bibnamefont
  {Feldman}}, \bibinfo {author} {\bibfnamefont {R.}~\bibnamefont {Austin}},
  \bibinfo {author} {\bibfnamefont {A.}~\bibnamefont {Dayem}}, \ and\ \bibinfo
  {author} {\bibfnamefont {E.}~\bibnamefont {Westerwick}},\ }\href@noop {}
  {\bibfield  {journal} {\bibinfo  {journal} {Applied Physics Letters}\
  }\textbf {\bibinfo {volume} {49}},\ \bibinfo {pages} {797} (\bibinfo {year}
  {1986})}\BibitemShut {NoStop}%
\bibitem [{\citenamefont {Feldman}\ \emph {et~al.}(1987)\citenamefont
  {Feldman}, \citenamefont {Austin}, \citenamefont {Fuoss}, \citenamefont
  {Dayem}, \citenamefont {Westerwick}, \citenamefont {Nakahara}, \citenamefont
  {Boone}, \citenamefont {Menendez}, \citenamefont {Pinczuk}, \citenamefont
  {Valladares} \emph {et~al.}}]{feldman1987}%
  \BibitemOpen
  \bibfield  {author} {\bibinfo {author} {\bibfnamefont {R.}~\bibnamefont
  {Feldman}}, \bibinfo {author} {\bibfnamefont {R.}~\bibnamefont {Austin}},
  \bibinfo {author} {\bibfnamefont {P.}~\bibnamefont {Fuoss}}, \bibinfo
  {author} {\bibfnamefont {A.}~\bibnamefont {Dayem}}, \bibinfo {author}
  {\bibfnamefont {E.}~\bibnamefont {Westerwick}}, \bibinfo {author}
  {\bibfnamefont {S.}~\bibnamefont {Nakahara}}, \bibinfo {author}
  {\bibfnamefont {T.}~\bibnamefont {Boone}}, \bibinfo {author} {\bibfnamefont
  {J.}~\bibnamefont {Menendez}}, \bibinfo {author} {\bibfnamefont
  {A.}~\bibnamefont {Pinczuk}}, \bibinfo {author} {\bibfnamefont
  {J.}~\bibnamefont {Valladares}},  \emph {et~al.},\ }\href@noop {} {\bibfield
  {journal} {\bibinfo  {journal} {Journal of Vacuum Science \& Technology B}\
  }\textbf {\bibinfo {volume} {5}},\ \bibinfo {pages} {690} (\bibinfo {year}
  {1987})}\BibitemShut {NoStop}%
\bibitem [{\citenamefont {Takemura}\ \emph {et~al.}(1992)\citenamefont
  {Takemura}, \citenamefont {Konagai}, \citenamefont {Nakanishi},\ and\
  \citenamefont {Takahashi}}]{takemura1992}%
  \BibitemOpen
  \bibfield  {author} {\bibinfo {author} {\bibfnamefont {Y.}~\bibnamefont
  {Takemura}}, \bibinfo {author} {\bibfnamefont {M.}~\bibnamefont {Konagai}},
  \bibinfo {author} {\bibfnamefont {H.}~\bibnamefont {Nakanishi}}, \ and\
  \bibinfo {author} {\bibfnamefont {K.}~\bibnamefont {Takahashi}},\ }\href@noop
  {} {\bibfield  {journal} {\bibinfo  {journal} {Journal of Crystal Growth}\
  }\textbf {\bibinfo {volume} {117}},\ \bibinfo {pages} {144} (\bibinfo {year}
  {1992})}\BibitemShut {NoStop}%
\bibitem [{\citenamefont {Dunstan}(1997)}]{dunstan1997}%
  \BibitemOpen
  \bibfield  {author} {\bibinfo {author} {\bibfnamefont {D.}~\bibnamefont
  {Dunstan}},\ }\href@noop {} {\bibfield  {journal} {\bibinfo  {journal}
  {Journal of Materials Science: Materials in Electronics}\ }\textbf {\bibinfo
  {volume} {8}},\ \bibinfo {pages} {337} (\bibinfo {year} {1997})}\BibitemShut
  {NoStop}%
\bibitem [{Note2()}]{Note2}%
  \BibitemOpen
  \bibinfo {note} {See supplementary online material for a brief derivation of
  Eq.~1.}\BibitemShut {Stop}%
\bibitem [{\citenamefont {Schenk}\ \emph {et~al.}(1996)\citenamefont {Schenk},
  \citenamefont {H{\"a}hnert}, \citenamefont {Duong},\ and\ \citenamefont
  {Niebsch}}]{Schenk1996}%
  \BibitemOpen
  \bibfield  {author} {\bibinfo {author} {\bibfnamefont {M.}~\bibnamefont
  {Schenk}}, \bibinfo {author} {\bibfnamefont {I.}~\bibnamefont {H{\"a}hnert}},
  \bibinfo {author} {\bibfnamefont {L.}~\bibnamefont {Duong}}, \ and\ \bibinfo
  {author} {\bibfnamefont {H.-H.}\ \bibnamefont {Niebsch}},\ }\href@noop {}
  {\bibfield  {journal} {\bibinfo  {journal} {Crystal Research and Technology}\
  }\textbf {\bibinfo {volume} {31}},\ \bibinfo {pages} {665} (\bibinfo {year}
  {1996})}\BibitemShut {NoStop}%
\bibitem [{\citenamefont {Andrusiv}\ \emph {et~al.}(1983)\citenamefont
  {Andrusiv}, \citenamefont {Grigorovich}, \citenamefont {Ilisavskii},\ and\
  \citenamefont {Ruvinskii}}]{Andrusiv1983}%
  \BibitemOpen
  \bibfield  {author} {\bibinfo {author} {\bibfnamefont {I.}~\bibnamefont
  {Andrusiv}}, \bibinfo {author} {\bibfnamefont {G.}~\bibnamefont
  {Grigorovich}}, \bibinfo {author} {\bibfnamefont {Y.}~\bibnamefont
  {Ilisavskii}}, \ and\ \bibinfo {author} {\bibfnamefont {M.}~\bibnamefont
  {Ruvinskii}},\ }\href@noop {} {\bibfield  {journal} {\bibinfo  {journal}
  {Sov. Phys. Solid State}\ }\textbf {\bibinfo {volume} {25}},\ \bibinfo
  {pages} {139} (\bibinfo {year} {1983})}\BibitemShut {NoStop}%
\bibitem [{Note3()}]{Note3}%
  \BibitemOpen
  \bibinfo {note} {See supplementary online material for additional remarks on
  the determination of the state of strain in the QW from HRXRD measurements
  and fits.}\BibitemShut {Stop}%
\bibitem [{\citenamefont {Novik}\ \emph {et~al.}(2005)\citenamefont {Novik},
  \citenamefont {Pfeuffer-Jeschke}, \citenamefont {Jungwirth}, \citenamefont
  {Latussek}, \citenamefont {Becker}, \citenamefont {Landwehr}, \citenamefont
  {Buhmann},\ and\ \citenamefont {Molenkamp}}]{Novik2005}%
  \BibitemOpen
  \bibfield  {author} {\bibinfo {author} {\bibfnamefont {E.~G.}\ \bibnamefont
  {Novik}}, \bibinfo {author} {\bibfnamefont {A.}~\bibnamefont
  {Pfeuffer-Jeschke}}, \bibinfo {author} {\bibfnamefont {T.}~\bibnamefont
  {Jungwirth}}, \bibinfo {author} {\bibfnamefont {V.}~\bibnamefont {Latussek}},
  \bibinfo {author} {\bibfnamefont {C.~R.}\ \bibnamefont {Becker}}, \bibinfo
  {author} {\bibfnamefont {G.}~\bibnamefont {Landwehr}}, \bibinfo {author}
  {\bibfnamefont {H.}~\bibnamefont {Buhmann}}, \ and\ \bibinfo {author}
  {\bibfnamefont {L.~W.}\ \bibnamefont {Molenkamp}},\ }\href@noop {} {\bibfield
   {journal} {\bibinfo  {journal} {Physical Review B}\ }\textbf {\bibinfo
  {volume} {72}},\ \bibinfo {pages} {035321} (\bibinfo {year}
  {2005})}\BibitemShut {NoStop}%
\bibitem [{\citenamefont {Ashcroft}\ and\ \citenamefont
  {Mermin}(1976)}]{Ashcroft1976}%
  \BibitemOpen
  \bibfield  {author} {\bibinfo {author} {\bibfnamefont {N.}~\bibnamefont
  {Ashcroft}}\ and\ \bibinfo {author} {\bibfnamefont {N.}~\bibnamefont
  {Mermin}},\ }\href@noop {} {\emph {\bibinfo {title} {{Solid State
  Physics}}}},\ HRW international editions\ (\bibinfo  {publisher} {Holt,
  Rinehart and Winston},\ \bibinfo {year} {1976})\BibitemShut {NoStop}%
\bibitem [{\citenamefont {Kvon}\ \emph {et~al.}(2008)\citenamefont {Kvon},
  \citenamefont {Olshanetsky}, \citenamefont {Kozlov}, \citenamefont
  {Mikhailov},\ and\ \citenamefont {Dvoretskii}}]{kvon2008}%
  \BibitemOpen
  \bibfield  {author} {\bibinfo {author} {\bibfnamefont {Z.~D.}\ \bibnamefont
  {Kvon}}, \bibinfo {author} {\bibfnamefont {E.}~\bibnamefont {Olshanetsky}},
  \bibinfo {author} {\bibfnamefont {D.~A.}\ \bibnamefont {Kozlov}}, \bibinfo
  {author} {\bibfnamefont {N.~N.}\ \bibnamefont {Mikhailov}}, \ and\ \bibinfo
  {author} {\bibfnamefont {S.~A.}\ \bibnamefont {Dvoretskii}},\ }\href@noop {}
  {\bibfield  {journal} {\bibinfo  {journal} {JETP Letters}\ }\textbf {\bibinfo
  {volume} {87}},\ \bibinfo {pages} {502} (\bibinfo {year} {2008})}\BibitemShut
  {NoStop}%
\bibitem [{\citenamefont {Olshanetsky}\ \emph {et~al.}(2014)\citenamefont
  {Olshanetsky}, \citenamefont {Kvon}, \citenamefont {Gerasimenko},
  \citenamefont {Prudkoglyad}, \citenamefont {Pudalov}, \citenamefont
  {Mikhailov},\ and\ \citenamefont {Dvoretsky}}]{olshanetsky2014}%
  \BibitemOpen
  \bibfield  {author} {\bibinfo {author} {\bibfnamefont {E.}~\bibnamefont
  {Olshanetsky}}, \bibinfo {author} {\bibfnamefont {Z.~D.}\ \bibnamefont
  {Kvon}}, \bibinfo {author} {\bibfnamefont {Y.~A.}\ \bibnamefont
  {Gerasimenko}}, \bibinfo {author} {\bibfnamefont {V.}~\bibnamefont
  {Prudkoglyad}}, \bibinfo {author} {\bibfnamefont {V.~M.}\ \bibnamefont
  {Pudalov}}, \bibinfo {author} {\bibfnamefont {N.~N.}\ \bibnamefont
  {Mikhailov}}, \ and\ \bibinfo {author} {\bibfnamefont {S.}~\bibnamefont
  {Dvoretsky}},\ }\href@noop {} {\bibfield  {journal} {\bibinfo  {journal}
  {JETP Letters}\ }\textbf {\bibinfo {volume} {98}},\ \bibinfo {pages} {843}
  (\bibinfo {year} {2014})}\BibitemShut {NoStop}%
\bibitem [{\citenamefont {Knap}\ \emph {et~al.}(2014)\citenamefont {Knap},
  \citenamefont {Sau}, \citenamefont {Halperin},\ and\ \citenamefont
  {Demler}}]{Knap2014}%
  \BibitemOpen
  \bibfield  {author} {\bibinfo {author} {\bibfnamefont {M.}~\bibnamefont
  {Knap}}, \bibinfo {author} {\bibfnamefont {J.~D.}\ \bibnamefont {Sau}},
  \bibinfo {author} {\bibfnamefont {B.~I.}\ \bibnamefont {Halperin}}, \ and\
  \bibinfo {author} {\bibfnamefont {E.}~\bibnamefont {Demler}},\ }\href@noop {}
  {\bibfield  {journal} {\bibinfo  {journal} {Physical Review Letters}\
  }\textbf {\bibinfo {volume} {113}},\ \bibinfo {pages} {186801} (\bibinfo
  {year} {2014})}\BibitemShut {NoStop}%
\bibitem [{\citenamefont {V{\"a}yrynen}\ \emph {et~al.}(2013)\citenamefont
  {V{\"a}yrynen}, \citenamefont {Goldstein},\ and\ \citenamefont
  {Glazman}}]{vayrynen2013}%
  \BibitemOpen
  \bibfield  {author} {\bibinfo {author} {\bibfnamefont {J.~I.}\ \bibnamefont
  {V{\"a}yrynen}}, \bibinfo {author} {\bibfnamefont {M.}~\bibnamefont
  {Goldstein}}, \ and\ \bibinfo {author} {\bibfnamefont {L.~I.}\ \bibnamefont
  {Glazman}},\ }\href@noop {} {\bibfield  {journal} {\bibinfo  {journal}
  {Physical Review Letters}\ }\textbf {\bibinfo {volume} {110}},\ \bibinfo
  {pages} {216402} (\bibinfo {year} {2013})}\BibitemShut {NoStop}%
\end{thebibliography}

\begin{thebibliography}{3}%
\makeatletter
\providecommand \@ifxundefined [1]{%
 \@ifx{#1\undefined}
}%
\providecommand \@ifnum [1]{%
 \ifnum #1\expandafter \@firstoftwo
 \else \expandafter \@secondoftwo
 \fi
}%
\providecommand \@ifx [1]{%
 \ifx #1\expandafter \@firstoftwo
 \else \expandafter \@secondoftwo
 \fi
}%
\providecommand \natexlab [1]{#1}%
\providecommand \enquote  [1]{``#1''}%
\providecommand \bibnamefont  [1]{#1}%
\providecommand \bibfnamefont [1]{#1}%
\providecommand \citenamefont [1]{#1}%
\providecommand \href@noop [0]{\@secondoftwo}%
\providecommand \href [0]{\begingroup \@sanitize@url \@href}%
\providecommand \@href[1]{\@@startlink{#1}\@@href}%
\providecommand \@@href[1]{\endgroup#1\@@endlink}%
\providecommand \@sanitize@url [0]{\catcode `\\12\catcode `\$12\catcode
  `\&12\catcode `\#12\catcode `\^12\catcode `\_12\catcode `\%12\relax}%
\providecommand \@@startlink[1]{}%
\providecommand \@@endlink[0]{}%
\providecommand \url  [0]{\begingroup\@sanitize@url \@url }%
\providecommand \@url [1]{\endgroup\@href {#1}{\urlprefix }}%
\providecommand \urlprefix  [0]{URL }%
\providecommand \Eprint [0]{\href }%
\providecommand \doibase [0]{http://dx.doi.org/}%
\providecommand \selectlanguage [0]{\@gobble}%
\providecommand \bibinfo  [0]{\@secondoftwo}%
\providecommand \bibfield  [0]{\@secondoftwo}%
\providecommand \translation [1]{[#1]}%
\providecommand \BibitemOpen [0]{}%
\providecommand \bibitemStop [0]{}%
\providecommand \bibitemNoStop [0]{.\EOS\space}%
\providecommand \EOS [0]{\spacefactor3000\relax}%
\providecommand \BibitemShut  [1]{\csname bibitem#1\endcsname}%
\let\auto@bib@innerbib\@empty
\bibitem [{\citenamefont {Dunstan}(1997)}]{dunstan1997}%
  \BibitemOpen
  \bibfield  {author} {\bibinfo {author} {\bibfnamefont {D.}~\bibnamefont
  {Dunstan}},\ }\href@noop {} {\bibfield  {journal} {\bibinfo  {journal}
  {Journal of Materials Science: Materials in Electronics}\ }\textbf {\bibinfo
  {volume} {8}},\ \bibinfo {pages} {337} (\bibinfo {year} {1997})}\BibitemShut
  {NoStop}%
\bibitem [{\citenamefont {Schenk}\ \emph {et~al.}(1996)\citenamefont {Schenk},
  \citenamefont {H{\"a}hnert}, \citenamefont {Duong},\ and\ \citenamefont
  {Niebsch}}]{Schenk1996}%
  \BibitemOpen
  \bibfield  {author} {\bibinfo {author} {\bibfnamefont {M.}~\bibnamefont
  {Schenk}}, \bibinfo {author} {\bibfnamefont {I.}~\bibnamefont {H{\"a}hnert}},
  \bibinfo {author} {\bibfnamefont {L.}~\bibnamefont {Duong}}, \ and\ \bibinfo
  {author} {\bibfnamefont {H.-H.}\ \bibnamefont {Niebsch}},\ }\href@noop {}
  {\bibfield  {journal} {\bibinfo  {journal} {Crystal Research and Technology}\
  }\textbf {\bibinfo {volume} {31}},\ \bibinfo {pages} {665} (\bibinfo {year}
  {1996})}\BibitemShut {NoStop}%
\bibitem [{\citenamefont {Andrusiv}\ \emph {et~al.}(1983)\citenamefont
  {Andrusiv}, \citenamefont {Grigorovich}, \citenamefont {Ilisavskii},\ and\
  \citenamefont {Ruvinskii}}]{Andrusiv1983}%
  \BibitemOpen
  \bibfield  {author} {\bibinfo {author} {\bibfnamefont {I.}~\bibnamefont
  {Andrusiv}}, \bibinfo {author} {\bibfnamefont {G.}~\bibnamefont
  {Grigorovich}}, \bibinfo {author} {\bibfnamefont {Y.}~\bibnamefont
  {Ilisavskii}}, \ and\ \bibinfo {author} {\bibfnamefont {M.}~\bibnamefont
  {Ruvinskii}},\ }\href@noop {} {\bibfield  {journal} {\bibinfo  {journal}
  {Sov. Phys. Solid State}\ }\textbf {\bibinfo {volume} {25}},\ \bibinfo
  {pages} {139} (\bibinfo {year} {1983})}\BibitemShut {NoStop}%
\end{thebibliography}
\end{document}